\documentclass[a4paper,11pt]{article}
\usepackage{inputenc}
\usepackage{amssymb,amsmath}
\usepackage[english]{babel}
\usepackage{epsfig,graphicx,color}

\newcommand{\be}{\begin{equation}\label}
\newcommand{\ee}{\end{equation}}
\newcommand{\prt}{\partial}
\newcommand{\p}{\prime}

\begin{document}

\begin{center}

{\bf \large ON A MODEL OF A CLASSICAL RELATIVISTIC PARTICLE \\
OF CONSTANT AND UNIVERSAL MASS AND SPIN}

\vskip4mm

{\bf V. Kassandrov$^{1,2}$,~ N. Markova$^{1,2}$,~ G. Sch\"{a}fer$^2$, ~A. Wipf$^2$}

\vskip2mm
$^1$ Institute of Gravitation and Cosmology,~Peoples' Friendship University of Russia, Moscow, Russia                

$^2$ Institute of Theoretical Physics, Friedrich-Schiller 
University, Jena, Germany

\end{center}

\noindent
{\bf Abstract}. {\footnotesize The deformation of the classical action 
for a point-like particle recently suggested by Staruszkiewicz 
gives rise to a spin structure which constrains the values of the 
invariant mass and the invariant spin  
to be the same for \emph{any solution} of the equations of motion.
Both these Casimir invariants, the square of the four-momentum vector
and the square of the Pauli-Luba\'{n}ski pseudo-vector, are shown to preserve the same fixed values
also in the presence of an arbitrary external electromagnetic field. 
In the ``free'' case, in the centre-of-mass reference frame, 
the particle moves along a circle of fixed radius with arbitrary 
varying frequency. In a homogeneous magnetic field, a number of rotational ``states'' 
is possible, with frequencies slightly different 
from the cyclotron frequency, and ``phase-like'' transitions 
with spin flops occure at some critical value of the particle's three-momentum. 
In the last section, the article of Kuzenko, Lyakhovich and Segal  
(1994) in which, in fact, an equivalent model had been earlier proposed and 
elaborated, is briefly reviewed and compared with Staruszkiewicz's approach and 
our results.}

\vskip2mm 


\section{Introduction. The Staruszkiewicz model}

From the times of Frenkel~\cite{Frenkel} and Mathisson~\cite{Matiss}, or from 
later works~\cite{BMT, Corben}, a 
lot of attempts have been undertaken to unambiguously formulate the dynamics of a
classical spinning particle~\cite{Nash}-\cite{Rylov}. Most of them deal
with generalizations of the classical point-mass Lagrangian $(-mc\sqrt{\dot{x}\dot{x}})$ through the introduction of terms with higher derivatives or
``inner'' variables, and then try to restrict the undesirable freedom by making use 
of some geometrical~\cite{Plyushchay,Nesterenko} or symmetry~\cite{Rivas,RivasDE} considerations. 
A thorough analysis of extra variables responsible for the spin 
structure has been carried out by Hanson and Regge~\cite{Hanson}. 
Rivas~\cite{RivasAH} proposed to make use of a complete set of 
parameters of the kinematical symmetry group and obtained considerable 
restrictions on the spin dynamics (so-called ``atomic hypothesis'').
 
However, a lot of problems arising within the classical description of 
spin still have to be solved. Corresponding Lagrangians, especially in 
their interaction part, are rather ambiguous, cumbersome and have no
correspondence with generally accepted gauge theories and with the structure 
of the Lorentz force in the spinless limit in particular~\cite{RivasInt}. The 
Schr\"odinger {\it zitterbewegung}  motion, being a common feature of different spin models, results in 
the problem of radiation of electromagnetic waves, and the invariant spin is 
not bound to preserve its value in the 
presence of external fields (as one expects for an innate characteristic of 
elementary particle). Thus, the ``zitterbewegung'' radius and the magnitude of 
spin are, as a rule, to be fixed ``by hands''. Generally, it is not quite  
clear which properties of the spin could arise in the framework of 
a successively classical relativistic model.

Meanwhile, in a short note~\cite{Starusz} Staruszkiewicz has offered a fairly simple relativistic model for a 
classical particle with spin~\footnote{Recently we became informed that 
nearly the same model has been suggested and elaborated in 1994 in the paper
~\cite{Kuzenko}. We present and discuss it in the last section (Afterword) 
below.}. Specifically, instead of the consideration of 
a general local reference frame related to the orientation of the spin vector, 
he restricted additional degrees of freedom by a {\it single null vector}  
$k_\mu(\tau), (k\cdot k)=0$ attached to each point of a trajectory of the classical particle. 
We note that Lorentz indices hereafter are labelled by Greek letters $\mu,\nu,...=0,1,2,3$, 
whereas $(a \cdot c )= a_0c_0 - \vec a\vec c$ denotes the Lorentz scalar product. 
The derivative $d/d\tau$ with respect to an invariant time parameter $\tau$ 
(identified later with the particle's proper time) is denoted by a dot.

Remarkably, if $l$ is a new universal constant, a {\it fundamental length}, 
corresponding to this additionally introduced variable, one essentially has
the only possibility to construct 
a dimensionless, relativistic and translation invariant quantity 
\be{zeta}
\zeta = l^2 \frac{(\dot{k}\cdot \dot{k})}{(k \cdot \dot  {x})^2}, 
\ee
which is moreover reparametrization $\tau\mapsto g(\tau)$ and 
projective $k_\mu \mapsto  h(\tau) k_\mu$ invariant, $g$ and $h$ 
being arbitrary smooth and monotonic (as to $g$) functions of the invariant  
time parameter $\tau$. 

Now one can write down the most general deformation of the well-known relativistic
invariant point-mass action described 
by a position four-vector $x_\mu(\tau)$ and a null direction four-vector $k_\mu(\tau)$ in a unique 
and simple form~\cite{Starusz}  
\be{act}
S=\int L d\tau = -mc\int d\tau \sqrt{\dot{x}\cdot\dot{x}}f(\zeta).
\ee
   
In order to fix a particular form of the unknown function $f(\zeta)$ 
~\footnote{For the energy to be positive definite one requires $f(\zeta)>0$ 
provided $m>0$}, Staruszkiewicz, inspired by the idea of ``irreducibility'' from 
the famous paper by E. Wigner~\cite{Wigner}, requires then for the values of 
both Casimir invariants of the Poincar\'e group to take one and 
the same value on any solution to the model (\ref{act}). He asserts that such a 
requirement leads to two differential equations for the unknown 
$f(\zeta)$. According to~\cite{Starusz}, these equations have only one common 
solution that uniquely defines the form of the function $f(\zeta )$ to be 
\be{func} 
f=\sqrt{1+\sqrt{-\zeta}}.
\ee

The two Casimir invariants associated with the action (\ref{act}) are the 
four-momentum vector square 
\be{I_1}
I_1 = (P\cdot P) 
\ee
and the square 
\be{I_2}
I_2 = (W\cdot W) 
\ee
of the Pauli-Luba\'{n}ski pseudovector
\be{luban}
W_{\mu}=-\frac{1}{2mc}\varepsilon_{\mu\nu\rho\sigma}M^{\nu\rho}P^{\sigma}, 
\ee
responsible for the invariant (i.e. independent on the origin) and 
conserved spin structure. In equation (\ref{luban}) $M^{\nu\rho}$ is the 
(conserved) total angular momentum tensor of the Lagrangian (\ref{act}). 
According to statements made in~\cite{Starusz}, for any solution 
of the equations of motion with the choice (\ref{func}) 
for the function $f(\zeta)$, the invariants $I_1$ and $I_2$ take the 
only possible values
\be{values}
I_1 = (mc)^2\quad\hbox{and}\quad I_2 = -\frac{1}{4}(mcl)^2.
\ee
Thus, in the framework of (generalized) classical mechanics of the 
``free'' particle one is able to elegantly describe its spin structure 
and, moreover, to ensure the property of spin (and mass) ``quantization''. 
For some preliminary chosen ``ethalon'' units $\{m, c, l\}$, the 
Staruszkiewicz model {\it corresponds to a point 
relativistic particle of constant and universal mass and spin}. In particular, one can
equate the fundamental length to the Compton wavelength $l=\hbar/mc$, 
so that the value of spin is then one half of the Planck constant, 

$$S=\sqrt{-W\cdot W}=\frac{1}{2}mcl =\frac{1}{2}\hbar .$$ 

In view of these intriguing properties, our main goal in this 
article is to prove the statements of Staruszkiewicz in~\cite{Starusz} 
and to generalize these to the case when the spinning particle
interacts with an external electromagnetic field. We shall also study
some unusual properties of exact solutions to the equations of motion
for a ``free'' particle and for a particle in a homogeneous magnetic field.

\section{Conservation laws and fixed invariants of the model}

Let us first note that due to the above mentioned local symmetry
$k_\mu\to h(\tau)k_\mu$ 
the quantity $\zeta$ in (\ref{zeta}) depends only on the {\it ratio} 
of the components of the null vector $k_\mu$, namely, on the components 
$n^a = k^a/k^0,~a=1,2,3$  of the unit three-vector $\vec n=\{n^a\},~~{\vec n}^2=1$. 
Specifically, one obtains
\be{zetaN}
-\zeta = l^2\frac{\dot{\vec n}\dot{\vec n}}{(\dot{x}_0-\vec n \dot{\vec x})^2}.
\ee
Thus, the individual components of the null vector $k_\mu$ are not
determined, and only  the {\it direction vector} $\vec n=\vec k / k^0$ will 
finally enter the equations of motion.  
However, to explicitly preserve relativistic invariance in the
variational principle an additional term $\lambda (k\cdot k)$ with Lagrangian multiplier $\lambda$ 
has to be added to the action (\ref{act}). 

We shall also from the very beginning switch on the electromagnetic
interaction by introducing the usual and (modulo total time derivative) 
gauge invariant  term $-(e/c)\dot{x}^\mu A_\mu$ with $e$ being the {\it electric charge} of the particle and 
$A_\mu$  the {\it four-potential} of the external electromagnetic field.
Thus, we shall deal with the action
\be{actN}
S=\int L d\tau,~~~L=-mc \sqrt{(\dot{x}\cdot \dot{x})}f(\zeta )-\frac{e}{c} (\dot{x} \cdot A)+\lambda (k\cdot k).
\ee

Consider now the integrals of motion related to translational 
and rotational symmetries of the model (\ref{actN}).  For the canonical
four-momentum $\pi_\mu=-\prt L/ \prt \dot{x}^\mu$ one obtains
\be{gen4mom}
\pi_\mu = P_\mu + \frac{e}{c} A_\mu, ~~~P_\mu := 
\frac{mc}{\epsilon}f\dot{x}_\mu + Bk_\mu,
\ee
where $\epsilon=\sqrt{\dot{x} \cdot \dot{x}}$, and 
\be{A}
B=-2mc\frac{\epsilon}{(k\cdot \dot{x})} \zeta f^\p(\zeta),
\ee 
with $\p$ denoting differentiation with respect to $\zeta$.    
Calculating the square of the {\it kinematical} 
four-momentum $P_\mu$, one obtains for the first Casimir 
invariant 
\be{I_1N}
I_1=(P\cdot P) = (mc)^2 (f^2-4\zeta f f^\p). 
\ee

Now one is ready to write down 
the first set of the Euler-Lagrange equations of motion
\be{eqmot1}
\dot{P}_\mu = mc\frac{d}{d\tau}\left(\frac{f}{\epsilon} \dot{x}_\mu - \frac{2 \epsilon \zeta f^\p}{(k\cdot \dot{x})} k_\mu\right) = 
F_\mu, 
\ee
where $F_\mu$ stands for the four-vector  of {\it Lorentz force},
\be{LorF}  
F_\mu = \frac{e}{c} F_{\mu\nu} \dot{x}^\nu, 
\ee
with $F_{\mu\nu}$ being the electromagnetic field strength tensor, 
$$F_{\mu\nu} = \prt_\mu A_\nu - \prt_\nu A_\mu,~~~\prt_\mu =: \frac
{\prt}{\prt x^\mu}.$$ 

The ``internal four-momentum'' $Q_\mu$ is a direct  
analogue of the ordinary one with respect to 
additional degrees of freedom related to the null four-vector $k_\mu$, 
\be{IntmomEM}
Q_\mu = -\frac{\prt L}{\prt \dot{k}^\mu} = G\dot{k}_\mu,~~~G=2mc\epsilon l^2 \frac{f^\p}{(k\cdot \dot{x})^2}, 
\ee
so that the second set of equations looks as follows:
\be{eqmot2}
\dot{Q}_\mu = -\frac{\prt L}{\prt k^\mu},~\rightarrow~\frac{d}{d\tau}(G\dot{k}_\mu)= B\dot{x}_\mu -2\lambda k_\mu,
\ee
where $B$ turns out to be the same as in (\ref{A}). 
We are now ready to introduce the {\it angular momentum} skew symmetric tensor 
\be{angularEM}
M_{\mu\nu} = x_{[\mu} P_{\nu]} + k_{[\mu} Q_{\nu]} = \frac{mc}{\epsilon} f x_{[\mu} \dot{x}_{\nu]} + B x_{[\mu} k_{\nu]} + G k_{[\mu} \dot{k}_{\nu]}, 
\ee
with the notation $A_{[\mu}B_{\nu]}= A_\mu B_\nu - A_\nu B_\mu$ to be used hereafter. 

The angular momentum is conserved in the absence of an external
field. Indeed, calculating with the help of
(\ref{gen4mom},\ref{eqmot1},\ref{IntmomEM},\ref{eqmot2})
its invariant time derivative, one obtains 
\be{angalt}
\dot{M}_{\mu\nu} =x_{[\mu}F_{\nu]} .
\ee

Let us now specify the form of the Pauli-Luba\'{n}ski pseudovector
(\ref{luban}) with the expression for kinematical part $P_\mu$ from 
(\ref{gen4mom}) to be used instead of the canonical four-momentum $\pi_\mu$. Making also 
use of (\ref{angularEM}), one obtains 
\be{lubanExp}
W_\mu = -\frac{G f(\zeta)}{\epsilon} \varepsilon_{\mu\nu\rho\sigma} k^\nu
\dot{k}^\rho \dot{x}^\sigma, 
\ee
so that the expression for the second Casimir invariant (\ref{I_2}) looks as follows:
\be{I_2N}
I_{2}=(W \cdot W)= \frac{G^2 f^2}{\epsilon^2} (k \cdot \dot{x})^2 
(\dot{k} \cdot {\dot k} ) = 4(mcl)^2 \zeta f^2 (f^\p)^2.  
\ee 

Now, following Staruszkiewicz and equating the two Casimir 
invariants (\ref{I_1N}) and (\ref{I_2N}) to their 
required values (\ref{values}), one arrives at two
differential equations for the same unknown $f(\zeta)$, namely, 
\be{eq1}
f^2-4\zeta f f^\p =1
\ee
and
\be{eq2}
-\zeta f^2 (f^\p)^2 = \frac{1}{16}. 
\ee

The first equation can easily be integrated leading (on account of
(\ref{zetaN}), $\zeta$ is negative on solutions)
to the following general solution:
\be{solgen}
f(\zeta) = \sqrt{1 + C \sqrt{-\zeta}},  
\ee
with $C$ being an arbitrary constant. For $C=0$, the model reduces to the 
canonical one for a free spinless particle. For nonzero $C\ne 0$, it is  
easy to verify that the function (\ref{solgen}) satisfies also  
the second equation (\ref{eq2}) provided $\vert C\vert =1$. 
Thus, one is finally left with {\it two different solutions} common to 
equations (\ref{eq1}) and(\ref{eq2})  
\be{solfunc}
f(\zeta):= f_\pm = \sqrt{1\pm \sqrt{-\zeta}},  
\ee
from which only the first one $f=f_+$ corresponds to that  
presented in~\cite{Starusz}. The second case $f=f_-$ 
must be studied on equal footing. Below we shall see that a distinction
between these two cases are essential and interesting from a physical
point of view. 

We see therefore that for the action (\ref{actN}) with fixed dimensional 
constants $m$, $l$ and $e$ and the function $f(\xi)$ 
taking one of the two obtained forms (\ref{solfunc}), expression (\ref{I_1N}) 
is identically a constant number; the same is true for expression (\ref{I_2N}). 
Thus, we have proved that both Casimir invariants take the same fixed
values for any solution of the Staruszkiewicz model {\it in the presence
of an arbitrary external electromagnetic field} provided the interaction 
is taken to be {\it minimal}. This is just the remarkable property one 
observes for a real quantum particle, whereas here it holds on a purely 
classical level. 
We are not aware of whether such a property holds in any other
model of classical spin dynamics.

Making now use of the expressions (\ref{eqmot1}) and (\ref{angalt}) for 
the variation of the linear and angular momentum, one obtains for the
derivative of the Pauli-Luba\'{n}ski pseudovector with respect to invariant 
time parameter
\be{lubanalt}
\dot{W}_\mu = -\frac{G}{2mc}\epsilon_{\mu\nu\rho\sigma}k^\nu\dot{k}^\rho F^\sigma. 
\ee
On account of the constant (and universal) values of the Casimir
invariants (\ref{I_1N}) and (\ref{I_2N}),
the relations $(P\cdot \dot{P})=0$ and $(W \cdot \dot{W})=0$ must
be fulfilled identically.
On the other hand, one can explicitly obtain these products using
(\ref{gen4mom}) and (\ref{angularEM})
together with the equations of motions (\ref{eqmot1}) and (\ref{angalt})
respectively.
Taking also into account the obvious identity 
$$(\dot{x}\cdot F) = F_{\mu\nu}\dot{x}^\mu\dot{x}^\nu \equiv 0,$$ 
one arrives at the following two relations:
\be{identCas}
0\equiv (P\cdot \dot{P})=-\frac{2mc^2\zeta f^\p}{(k\cdot \dot{x})}(k\cdot F),~~~
0\equiv (W \cdot \dot{W})=-\frac{mc^2l^2}{4(k\cdot \dot{x})}(k\cdot F),
\ee
which should be fulfilled ``on shell''. 
Thus, for any solution to the equations of motion {\it the null four-vector 
$k_\mu$ is necessarily orthogonal to the 
four-vector $F_\mu$ of the Lorentz force},  
\be{kF}
(k\cdot F)=F_{\mu\nu}k^\mu \dot{x}^\nu \equiv 0
\ee
Of course, the property (\ref{kF}) can be derived explicitly from the 
equations of motion themselves, without any direct account of the 
identities (\ref{I_1N}) and (\ref{I_2N}); 
however, this  requires rather lengthy computations. From the 
condition (\ref{kF}) it also follows that the orientation of vector $\vec n$ 
is instantaneously in correspondence with the direction of electromagnetic 
field and, even at the initial instant, cannot be set in an arbitrary manner. 
Below (section 4) we shall consider this 
property in the particular case of a homogeneous magnetic field.

Let us now come back to the whole system of equations of motion 
(\ref{eqmot1}),(\ref{eqmot2}) and identify hereafter the parameter $\tau$ 
with proper time along the particle's trajectory so that one has 
$$\epsilon=c \sqrt{(u\cdot u)}=c,$$ 
with $u_\mu=\dot{x}_\mu/c$ being the four-velocity unit vector. 
Eliminating also $\lambda$ and passing 
thus to the unit three-vector $n^a=k^a / k^0$, one obtains from (\ref{eqmot2}):
\be{eqmot2N}
\frac{l^2}{c^2}\frac{d}{d\tau}\left(\frac{f^2 f^\p \dot{\vec n}}{Z^2}\right)=
\frac{\zeta f f^\p}{Z} (u_0{\vec n} -{\vec u}), 
\ee
where the abbreviation
\be{Z}
Z:=f(u_0-{\vec n} {\vec u})\equiv f (k\cdot u)/k_0
\ee
was introduced.  Separating now the spatial part in (\ref{eqmot1}), 
we close the system of equations of motion as follows:
\be{eqmot1S}
\frac{d}{d\tau}\left(f{\vec u} - \frac{2\zeta f f^\p}{Z} \vec n\right) =
\frac{e}{mc}(u_0 \vec E + {\vec u}\times \vec H), 
\ee
$\vec E, \vec H$ being the electric and magnetic field strength 
vectors respectively. In addition, the remaining temporal part of (\ref{eqmot1}) reads
\be{eqmot1T}
\frac{d}{d\tau}\left(f u_0 - \frac{2 \zeta f f^\p}{Z}\right) =
\frac{e}{mc}\vec E \vec u, 
\ee
and is not independent but follows from (\ref{eqmot1S}): it just represents 
the theorem for the change of energy.

In general, the equations for the four-momentum (\ref{eqmot1S}), (\ref{eqmot1T}) differ 
in structure from the canonical ones but reduce to the latter 
in the limit $l \rightarrow 0$. Later on we shall see if such reduction is in 
agreement with the additional equations (\ref{eqmot2N}) for the unit 
three-vector $\vec n$.

\section{``Centre-of-mass frame'' and dynamics of ``free'' spinning particle}

Now we switch off the external electromagnetic field by
setting $F_\mu=0$ in order to see if some internal motion
(``zitterbewegung'') of the ``charge-position vector'' relative to 
the centre-of-mass can be accomplished 
by a ``free'' spinning particle, in analogy with many other  
classical spin models (see, e.g., the review in~\cite{Rivas}). To start with, we can take
into account the conservation of momentum (\ref{gen4mom}) and 
angular momentum (\ref{angularEM}) in absence of a field.  
This allows one to pass, through a Lorentz boost and a three-translation, 
to a distinguished {\it centre-of-mass} reference frame defined by the 
following conditions:
\be{cenmass}
P_a = 0~~~(P_0 = mc),~~~M_{0a}=0, ~~~a=1,2,3. 
\ee 
Equations (\ref{eqmot1S}),(\ref{eqmot1T}) now simplify to  
$$\vec u - \frac{2\zeta f^\p}{Z}\vec n =0,$$ 
and 
$$u_0 - \frac{2\zeta f^\p}{Z} =\frac{1}{f},$$
where $Z$ is defined by (\ref{Z}).
Multiplying the first of these equations by $\vec n$ and taking in account 
the second one, one gets $Z=1$.  Now, after substitution of the particular form (\ref{solfunc}) 
of $f(\zeta)$,  the first set of equations of motion takes the simple form 
\be{fundN}
\vec u = \pm u \vec n, ~~~u=\vert \vec u \vert = \frac{\sqrt{-\zeta}}{2 f} = 
\frac{\omega_0 l}{2 c},
\ee
and 
\be{fund0N}
u_0 = \frac{1}{f} \pm u,.   
\ee
where the characteristic ``frequency'' $\omega_0$, depending 
on $\zeta$, comes into play, 
\be{omega}
\omega_0:= \frac{c\sqrt{-\zeta}}{lf(\zeta)}, 
\ee 
and the different signs correspond to those of the functions $f=f_\pm$ 
in (\ref{solfunc}). Note that the direction three-vector $\vec n$ is, respectively, parallel or antiparallel 
to the three-velocity,  and that equation (\ref{fund0N}) is a direct consequence of 
(\ref{fundN}) as long as the function $f(\zeta)$ satisfies the defining relation (\ref{eq1}).  

The second set of equations (\ref{eqmot2N}) simplifies now to
\be{eqdirf}
\frac{d}{d\tau}\left(\frac{\dot{\vec n}}{\omega_0}\right) = -\omega_0 \vec n,
\ee
and together with (\ref{fundN}) forms a closed system defining the position 
vector $\vec x$ and the direction unit vector $\vec n$. Taking vector product 
of the above equations with $\vec n$, one finds that for any solution there exists 
a {\it conserved} pseudovector~\footnote{This is in fact equal to the 
spin vector $\vec S$, see equation (\ref{LubanComp}).}
\be{psvector}
{\vec \nu} = \frac{1}{\omega_0}\vec n \times \dot{\vec n},~~~\dot{\vec \nu} = 0.
\ee
Since this is {\it constant} and orthogonal to both $\dot{\vec n}$ and 
${\vec n}$, the latter at all times lies in a fixed plane. From the parallelism
(\ref{fundN}) of ${\vec n}$ and the three-velocity vector ${\vec u}$ one 
concludes that the spatial motion of the particle is also   
in a {\it plane}.

As result, for the unit vector ${\vec n}$ the most general 
(up to a 3D rotation and time shift) form is as follows:
\be{vecN}
\vec n =\{\cos{\kappa}, \sin{\kappa}, 0 \}, 
\ee 
with $\kappa=\kappa(\tau)$ being some function of the proper time parameter. 
Now  equations (\ref{eqdirf}) immediately lead to the relation
\be{rel}
\dot{\kappa} = \pm \omega_0, 
\ee
whence the dependence of $\omega_0$ on time remains
undetermined and different signs correspond 
to opposite directions of rotation (in definitely oriented  
centre-of-mass reference frame).

\begin{figure}[!ht]
\begin{center}
\includegraphics[scale=.8]{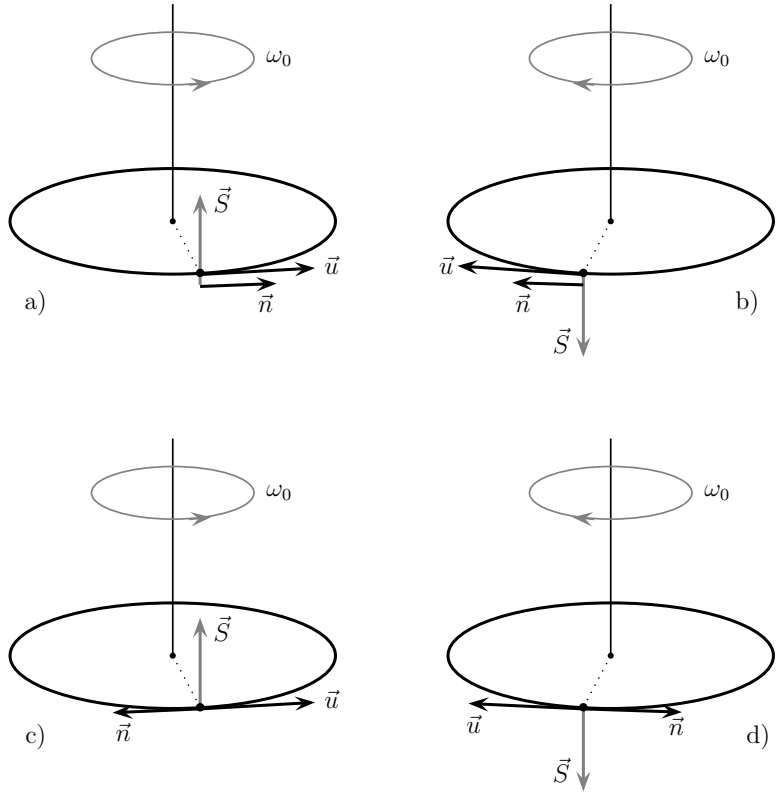}
\caption{\label{spin1} 
Relative orientations of the three-velocity $\vec u$, 
the direction three-vector $\vec n$ and the spin three-vector $\vec S$ 
under uniform internal rotation in the ``free'' case and in  
the centre-of-mass reference frame with definite orientation.}
\end{center}
\end{figure}

For both, 
the relative orientations of the three-vectors $\vec n$ and $\vec u$, in accord with 
(\ref{fundN}), may be different. However, the direction of the spin 
three-vector $\vec S = \{W^a\}$ is the same for both orientations 
and depends only on the direction of rotation. Indeed, 
from the general expression (\ref{lubanExp}) one has 
\be{LubanComp}
\vec S = \{W^a\}=2ml^2f^\p f^2 \vec n \times \dot {\vec n} = 
\pm 2 mcl (ff^\p \sqrt{-\zeta}) \vec e_z = \pm \frac{1}{2}mcl \vec e_z, 
\ee
where $\vec e_z$ is a unit vector along the $z$-axis, and the expression in 
parentheses is taken to be $1/4$ in accordance with the second defining equation 
(\ref{eq2}) for the generating function $f(\zeta)$. Thus, the spin vector is orthogonal 
to the plane of rotation, and its sign depends only on the 
direction of rotation. As a whole, four different {\it degenerate} 
``configurations'' are possible (Fig.\ref{spin1})~\footnote{Physical  
situations represented by a) and b), as well as by c) and d) panels 
at Fig.1, are in fact equivalent but can be distinguished by a particular 
observer.}. As for the temporal 
component $W_0$ of  the Pauli-Luba\'{n}ski four-vector, it is
proportional to $(\dot{\vec n} \cdot \vec n \times \vec u)$ and, therefore, zero. 
For the position vector one obtains then by integrating 
the equations (\ref{fundN})
\be{position}
\vec x = \frac{l}{2}\{\sin{\kappa}, -\cos{\kappa},0 \},
\ee
where the constant of integration was set to zero since the origin is 
placed at the centre-of-mass point (\ref{cenmass}), $M_{0a}=0$.   
We see  that {\it the ``free'' particle always moves along a circle with fixed radius}
\be{radius}
R=\frac{l}{2}, 
\ee 
and when the value of the spin is set equal to one half of the Planck's constant, 
$S=\hbar/2$, then {\it the diameter of the circle of rotation is exactly equal 
to the Compton length}. Thus, the particle itself is not located at its 
own center of mass but circles around the latter. This is a common feature of 
different classical spin models~\cite{Rivas};  however, the radius therein is 
usually quite arbitrary and is fixed (``quantized'') {\it ad hoc}.

Moreover, only in the considered model one comes across the 
stochastic-type {\it zittebewegung} phenomenon in full sense of the word. Indeed, 
the rotation frequency (\ref{omega}) defined according to (\ref{zetaN}) 
through the $\zeta$-variable, remains
indefinite or, uncontrolled. Indeed, calculating $\zeta$ from  
the defining formula (\ref{zetaN}) and making use of 
the expressions (\ref{vecN}),(\ref{rel}) and (\ref{position}) for 
the corresponding variables, one obtains an identity. Thus, the 
time variation of the parameter $\zeta$ and frequency $\omega_0$ 
can indeed be arbitrary. 

It is noteworthy to point out that independently on the particular form of the considered 
model, the simple equation (\ref{eqdirf}) for the unit three-vector ${\vec n}$ turns 
to describe, in account of its general solution (\ref{vecN}), an important 
dynamical system, namely, {\it a plane rotator with arbitrary varying 
frequency} $\omega(\tau)$. It is also of interest that
introducing a new ``internal'' time scale $\sigma(\tau)$ 
(more precisely, {\it phase} scale), 
\be{Tscale}
d\sigma = \omega_0(\tau) d\tau
\ee
one reduces (\ref{eqdirf}) to the equations 
\be{nss}
\frac{d^2\vec n}{d\sigma^2} = -\vec n,
\ee
with an unique solution corresponding to 
a plane rotation (clock- or anticlockwise) with {\it constant} unit frequency 
(in the specially adopted time scale). 
Of course, the trajectory of the particle, a circle of 
fixed radius (\ref{radius}), 
is itself invariant under any time scale reparametrizations.  
On the other hand, we shall see below that in realistic situations, 
when a constant external field is present, the frequency of rotation is as a rule  
constant and fixed in value.

In the simplest case one can consider the value of $\sqrt{-\zeta}$ to be a 
constant, with its range for the choice $f=f_-$ of the function (\ref{solfunc}) 
being 
restricted by the requirement $0 <-\zeta < 1$.  Then, according to (\ref{vecN}),(\ref{rel}), our ``free'' particle
accomplishes a {\it uniform rotation} with constant in time (but 
arbitrary!) frequency $\omega_0=c\sqrt{-\zeta}/l f(\zeta)$ (the direction of 
rotation can be arbitrary whereas the relative orientations of 
the three-vectors $\vec u$ and $\vec n$ in view of (\ref{fundN}) 
are different for the two different generating functions $f= f_\pm$).

For the temporal component of the four-velocity one gets then: 
\be{vel_0}
u_0 = \sqrt{1+u^2} = \sqrt{1+\left(\frac{\omega_0 l}{2 c}\right)^2}.
\ee
One can easily check that the same expression for $u_0$ 
follows from (\ref{fund0N}). This allows one to pass from the proper 
time parameter $\tau$ to the laboratory time, 
\be{time}
d\tau=cdt/u_0,
\ee
and to obtain the three-velocity $v^a = d{x}^a/dt$ and 
physical frequency of rotation $\omega$. One finds
\be{3velf}
\vec v = \frac{\omega l}{2} \{\cos{\omega t}, \sin{\omega t}, 0\}, ~~~   
\omega:=\frac{\omega_0}{\sqrt{1+(\omega_0 l/2c)^2}}.
\ee
One can see that the linear velocity never exceeds the 
speed of light, $v^2= \vec v \vec v < c^2$. It is also 
interesting (and even striking!) that the spin vector does not 
depend on the speed of rotation of the particle.
In particular, the frequency $\omega_0$ (and $\omega$) 
can be arbitrarily small and close to zero, so that in the limit one has an  
indefinite direction of the spin vector but its modulus remains 
fixed.  

Mathematically, this corresponds to an uncertainty of the type $0/0$ in the 
expression for the second Casimir invariant (\ref{I_2N}). When the
sign of frequency and direction of rotation are changed, one has 
a {\it spin-flop transition} at this critical point. In the 
next section we shall see that the same phenomenon also takes place 
for the particle in a homogeneous magnetic field and thus may be of 
universal nature. In the ``free'' case, however, {\it the law of motion
may be arbitrary}. Particularly, when $\zeta$ and thus $\omega$ are  
constant, the position vector will move (as it usually takes place 
in the classical spin models) along a {\it helix-like} curve.

\section{Frequency shift and spin-flop transitions in magnetic field}
Here we consider the classical problem of a relativistic charged particle in a {\it constant and 
homogeneous magnetic field}. The action (\ref{act}) of the
spinning particle contains the coupling of the particle to the
magnetic field and one of the two generating functions $f=f_\pm$ 
in (\ref{solfunc}). The energy ${\cal E}$ is then conserved,  so 
that instead of (\ref{eqmot1T}) one gets
\be{energy}
f(u_0 - \frac{y}{Z}) = \frac{\cal E}{mc^2}= \delta = \hbox{const} >1,
\ee
whereas the principal equation (\ref{eqmot1S}) for three-momentum $\vec P$ 
takes the form
\be{magnP}
\frac{d}{d\tau}\left(\frac{\vec P}{mc}\right) = \frac{d}{d\tau}\left(f\vec u - f\frac{y}{Z}\vec n\right) = -\Omega \vec u \times \vec h.
\ee 
Here $Z=f(u_0-\vec u \vec n)$ has been defined in (\ref{Z}),
$y$ is given by
\be{y}
y = \pm \frac{\sqrt{-\zeta}}{2 f_{\pm}}=\pm\frac{\omega_0 l}{2c},
\ee
and $\Omega$ stands for the canonical {\it cyclotron frequency} 
(with respect to the proper time scale) of a particle 
in a homogeneous magnetic field of strength $H$, 
\be{Omega}
\Omega = \frac{e H}{m c}. 
\ee 
Here we assume that the charge of the particle is elementary and {\it negative} (electron), 
whereas $\vec h$ is the unit vector in the direction of the magnetic field which is
aligned with the $z$-axis.

It follows from (\ref{magnP}) that the formerly considered ``free'' case,  
corresponding to $H=0$ and ${\vec P}=0$,
is not a well defined limit of the motion in a homogeneous magnetic field.
The limiting ``free state'' resulting from the limiting process will 
actually depend on the {\it ratio} of the external parameters $\Omega$ and
${\cal P} = \vert {\vec P} \vert$ as the two approach zero, 
and this points to the degeneracy of the ``free state".

The component of the three-momentum along $\vec h$ is conserved, so 
that we can get rid of the corresponding uniform motion by setting 
$P_z = 0$. However, for the spinning particle this condition does not imply 
$u_z =0$; instead one gets $u_z = (y/Z)n_z$.

The resulting motion in $z$-direction could correspond to a
{\it precession} of the spin vector. Here, however, 
we shall  only consider the simplest case of a {\it plane} motion 
and hence set $u_z = n_z = 0$. 

From the identity (\ref{kF}) for the case of a pure 
magnetic field it follows
\be{plane}
\vec h \cdot (\vec u \times \vec n)= 0, 
\ee
so that the vectors $\vec{h},~\vec{u}$, and $\vec n$ {\it lie instantaneously in 
one and the same plane}. For the considered motion in the plane orthogonal to magnetic field this means 
that the vectors $\vec u$  and $\vec n$ are either parallel or antiparallel, that is, 
\be{parallel}
\vec u = (\vec u \vec n) \vec n.
\ee

Now the equations (\ref{eqmot2N}) for the unit vector $\vec n$ 
can be simplified to the form 
\be{magnN}
\frac{d}{d\tau}\left(  \frac{\dot{\vec n}}{\omega_0 Z^2}\right) = -\omega_0 \vec n,
\ee
strongly resembling the evolution equation in the ``free'' case (\ref{eqdirf}). 

Contracting the equations (\ref{magnP}) with $\vec n$, one 
easily gets
\be{piO}
f(\vec u \vec n - \frac{y}{Z})= \pm \frac 
{\cal P}{mc} = \gamma = \hbox{const} ,
\ee
where $\cal P$ stands for the conserved {\it modulus} of the
three-momentum. By virtue of the fixed mass invariant one 
has then the usual energy-momentum relation
\be{energymom}
\delta = \sqrt{1+\gamma^2},
\ee
where $\gamma$ may be either positive or negative.
In fact, it is the projection of three-momentum onto the $\vec n$-direction 
whose sign depends on relative orientation of $\vec n$ and $\vec u$, see below.

Combining now equations (\ref{energy}) and (\ref{magnP}), one   
finds
\be{ZN}
Z=\delta-\gamma = \hbox{const} >0,
\ee
so that the considered system of equations simplifies to
\be{piON}
\gamma \dot{\vec n} = -(\vec u \vec n)\Omega\, \vec n \times \vec h,
\ee
\be{magnNN}
\frac{d}{d\tau}\left( \frac{\dot{\vec n}}{\omega}\right) = -\omega \vec n,~~~ 
\omega =  \omega_0 (\delta - \gamma) \equiv  \frac{\sqrt{-\zeta} c}
{f l}(\delta - \gamma), 
\ee  
and two additional (though not independent) relations
\be{rel0}
\delta = fu_0 - \frac{yf}{\delta - \gamma},
\ee
\be{rel1}
\gamma = f(\vec u \vec n) - \frac{yf}{\delta - \gamma}.
\ee

Equations (\ref{magnNN}) may be solved similarly as in the 
``free'' case leading to the solution
\be{anzatz}
\vec n = \{\cos \kappa, \sin \kappa, 0\}, ~~~\dot{\kappa}=\pm \omega.
\ee
Substituting this into the first set of equations (\ref{piON}), one 
obtains
\be{main}
\dot{\kappa}=\Omega \frac{(\vec u \vec n)}{\gamma},
\ee
and, because  of (\ref{anzatz}), finds for the {\it rotational frequency}
\be{freq}
\omega = \Omega \frac{u}{\vert \gamma \vert}.  
\ee
For the {\it radius} $R$ of the circular orbit one obtains the same expression   
\be{radiusM}
R =\frac{c u}{\omega} = \frac{c\vert \gamma \vert}{\Omega},     
\ee
as for a canonical spinless particle. Analogously, 
the expression (\ref{LubanComp}) for the spin vector 
\be{spinM}
\vec S = \{W^a\}= 
\pm \frac{1}{2} mcl \vec e_z
\ee
is still valid in  the considered case of a {\it plane} motion in a magnetic field. 
We conclude that the spin is always {\it polarized} along the $z$-direction, 
with its sign being dependent on the direction of rotation (which
may differ from the canonical one, see below).

From (\ref{rel1}) one concludes now that the {\it frequency of rotation 
$\omega$ does not coincide with the canonical cyclotron frequency 
$\Omega$},
\be{freqratio}
\frac{\omega}{\Omega} = \frac{u}{\vert\gamma\vert}\ne 1.
\ee
Note that both frequencies are measured here with respect to 
{\it proper} time of the particle; however, the ratio (\ref{freqratio}) 
is invariant under any change of time scale. 

Substituting the expressions for $(\vec u \vec n)$ from  
(\ref{rel1}) and for $\omega$ from (\ref{magnNN}) one arrives at the 
following exact equation for the value of the ``hidden parameter'' 
$\sqrt{-\zeta}$ :
\be{eqzeta}
\sqrt{-\zeta}\left(\pm \frac{\delta-\gamma}{\alpha}\mp \frac{1}
{2\gamma (\delta- \gamma)}\right) =1,
\ee
where the following {\it small parameter} $\alpha$ appears:
\be{alpha}
\alpha = \frac{\Omega l}{c} = \frac{\hbar \Omega}{mc^2}.
\ee   
To get the last relation, we assumed $l=\hbar/mc$ so that  
the value of spin would be equal to that of a quantum mechanical 
electron projected onto the $z$-axis. Note that all signs in 
(\ref{eqzeta}) are independent, the left pair corresponds 
to opposite directions of rotation and the right pair -- to the
possible choices $f_\pm$ for the generating function (\ref{solfunc}). 
In addition one should take into account that the sign of $\gamma$ is 
not determined. Later on we shall consider various possible 
combinations of these signs in more detail. 

Observe that for the particular case of an {\it electron}, even in the
strongest magnetic fields achieved in laboratories ($H\sim 10^4$\,Gauss),  
the numerical value of $\alpha$ does not exceed $10^{-15}$. Moreover, 
another dimensionless parameter $\alpha/\vert\gamma\vert$ seems to be of 
fundamental importance in  the model. In fact, a
{\it macroscopic} radius of rotation 
$R=c\vert\gamma\vert / \Omega$ greatly exceeds the Compton wavelength 
$l$,~~$R\gg l$, so that for the above parameter one gets, 
\be{algam}
\frac{\alpha}{\vert\gamma\vert} \ll 1.
\ee
Thus, the two external quantities governing the classical behavior of 
the electron in a homogeneous magnetic field, namely, the strength 
$H\sim \alpha$ of the field and the three-momentum ${\cal P}\sim \vert\gamma\vert$, 
define two small parameters. We shall see that the latters determine small 
{\it corrections} to the rotational characteristics of a spinning particle 
in a homogeneous magnetic field. 

\begin{figure}[!ht]
\begin{center}
\includegraphics[scale=.8]{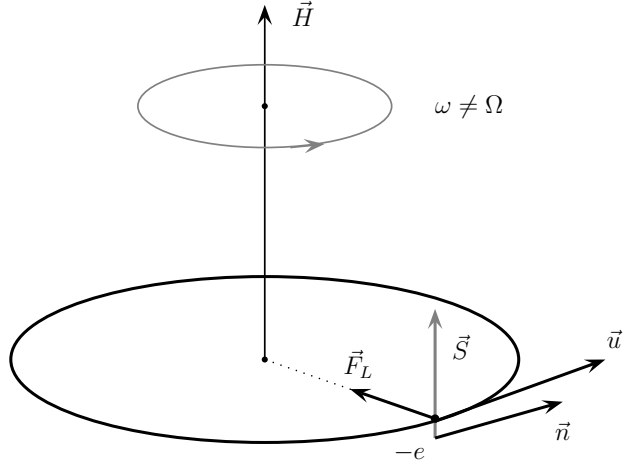}
\caption{\label{spin2} 
``Canonical'' orientation of the three-velocity $\vec u$, magnetic field 
$\vec H$ and Lorentz force $\vec F_L$ in the ``macroscopic'' case 
$R \gg l$.}
\end{center}
\end{figure}

For such a {\it macroscopic} motion, the direction of rotation is 
defined by the orientation of the ordinary Lorentz force towards the centre of 
rotation, see Fig.\ref{spin2},
The canonical motion corresponds to a positive sign in the expression 
for the frequency (\ref{anzatz}), so that from the principal equation
(\ref{eqzeta}) one gets for $\zeta$ the following expression: 
\be{eqzetaN}
\sqrt{-\zeta} = \frac{\alpha}{\delta-\gamma}\left(1\mp \frac{(\alpha/\gamma)}{2(\delta-\gamma)^2}\right)^{-1},
\ee
and this yields then the three-velocity and frequency of rotation, 
\be{correct}
\frac{u}{\vert\gamma\vert}= \frac{\omega}{\Omega}= 
\frac{(\delta-\gamma)\sqrt{-\zeta}}{\alpha f_{\pm}} =  
\frac{1}{f_{\pm}}\left(1\mp \frac{(\alpha/\gamma)}
{2(\delta-\gamma)^2}\right)^{-1},
\ee
where  the sign in parentheses correlates with the choice of 
generating function $f_\pm$. 

The last expression is exact and general, provided the direction 
of rotation is the canonical one. It describes {\it four different configurations} 
corresponding to the choice of $f=f_\pm$ and the different orientations of 
$\vec n$ relative to $\vec u$. They are all non-degenerate and define different 
three-velocities and frequencies for the same total three-momentum $\vert \gamma \vert$.  
In particular, for the relative {\it frequency shift} 
$$\Delta=(\omega-\Omega)/\Omega\ll 1$$
one obtains in leading order in the small parameters 
$\alpha$ and $\alpha/\vert\gamma\vert$:
\begin{eqnarray}
a)~~f=f_+,\quad -\vec n \parallel \vec u~~(\gamma < 0):&&\Delta \approx   
-\frac{(\alpha/\vert \gamma\vert)}{2(\delta+\vert\gamma\vert)^2}-\frac{\alpha}
{2(\delta+\vert\gamma\vert)};~~
\Delta < 0\label{shift+bar}
\label{shift-par}\\
b)~~f=f_-,\quad\;\;\;\vec n \parallel \vec u~~(\gamma > 0):&&\Delta \approx   
-\frac{(\alpha/\gamma)}{2(\delta-\gamma)^2}+\frac{\alpha}{2(\delta-\gamma)};
\label{shift+par}\\
c)~~f=f_+,\quad\;\;\;\vec n \parallel \vec u~~(\gamma > 0):&&\Delta \approx   
\frac{(\alpha/\gamma)}{2(\delta-\gamma)^2}-\frac{\alpha}{2(\delta-\gamma)};
\label{shift+anti}\\
d)~~f=f_-,\quad-\vec n \parallel \vec u~~(\gamma < 0):&&\Delta \approx  
\frac{(\alpha/\vert \gamma\vert)}{2(\delta+\vert\gamma\vert)^2}+\frac{\alpha}
{2(\delta+\vert\gamma\vert)};~~\Delta >0.\label{shift-anti}
\end{eqnarray}
Note that for the configurations with negative $\gamma$ the sign of the frequency shift is 
fixed; this is easy to understand taking in account that 
the addition to the canonical expression for the three-momentum in (\ref{magnP}) is
positive for configuration a) and negative for configuration d) so that the period 
of rotation has to increase or decrease, respectively. In the other two situations 
the effect depends on the relative magnitudes of magnetic field and 
three-momentum. 

In the {\it ultra-relativistic} case ${\cal E}/mc^2=
\delta \gg 1$, the parameter $(\delta - \gamma)$, for 
{\it configurations with parallel $\vec n$ and $\vec u$ 
and~ $\gamma\sim 
\delta >0$}, decreases as $1/2\delta$, and one gets for 
(\ref{shift+anti}) and (\ref{shift+par}) respectively:
\be{ultra}
f=f_\pm,~~~\vec n \parallel \vec u:~~~\Delta \approx \pm \alpha \delta 
\equiv \pm \left(\frac{\hbar \Omega}{mc^2}\right)
\left(\frac{\cal E}{mc^2}\right). 
\ee 

Even in this case, however, the frequency shift is hardly detectable in
a laboratory. Indeed, for achievable field strengths of
$10^4$\,Gauss, the parameter $\alpha$ is approximately $10^{-15}$ for electrons, 
whereas for an energy of $100$\,GeV one gets 
$\delta \sim 2\cdot 10^5$ such that the relative frequency shift is less
than $\Delta \sim 10^{-10}$. This seems to be too small to be
detectable on account  of the required constancy of energy and 
homogeneity of the magnetic field. We shall, however, postpone the 
detailed discussion of this problem to a forthcoming publication. 

Let us now turn to the more complicated and  questionable non-relativistic
case of  small three-momenta $\vert \gamma \vert$. In the model, this 
corresponds to a {\it microscopic} radius 
of the electron's gyration $R\sim l = \hbar/mc\sim 2.5\cdot 10^{-12}m$ 
and such microscopic orbits, as generally accepted, should be
described by quantum theory. 
Nonetheless, we shall now consider solutions corresponding to the 
microscopic orbits in a magnetic field from a purely classical point of view.

For small three-momenta one can set $\delta - \gamma \approx 1$
in the general formula (\ref{eqzeta}) and obtain for the $\sqrt{-\zeta}$ parameter
\be{microzeta}
\sqrt{-\zeta} \simeq \alpha (\pm 1 \mp \frac{\alpha}{2\gamma})^{-1}, 
\ee
and for the frequency of gyration
\be{omegamicro}
\frac{\omega}{\Omega} \simeq \frac{1}{f_\pm}(\pm 1 \mp \frac{\alpha}
{2\gamma})^{-1},
\ee
assuming for the expression in parentheses to be non-negative on solutions.   

\begin{figure}[!ht]
\begin{center}
\includegraphics[scale=.9]{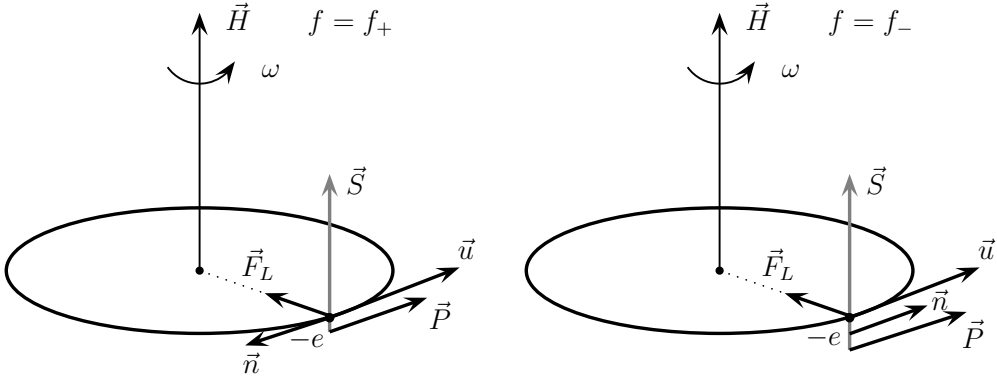}
\caption{\label{spin3} 
Configurations for the spinning particle rotating in magnetic field that
exist for any value of the particle's momentum. On the left a) $f=f_+$
and on the right b) $f=f_-$.}
\end{center}
\end{figure}

In particular, for $\vert \gamma \vert > \alpha/2$, the parameter $\sqrt{-\zeta}$  
is only positive for the canonical direction of gyration (this 
corresponds to the positive sign for the left pair in parentheses). 
In this case, the two solutions for which 
$$\sqrt{-\zeta} \simeq \alpha(1+\alpha/2\vert \gamma \vert)^{-1}$$ 
are well defined even for very small values of three-momentum $\vert \gamma \vert$. 
They correspond to the following configurations: 
\begin{eqnarray*}
a)\quad f=f_+&,&\gamma<0\quad(-\vec n \parallel \vec u)\\
b)\quad f=f_-&,&\gamma>0\quad\;\;(\vec n \parallel \vec u),
\end{eqnarray*}
see Fig.\ref{spin3}.
For $\vert \gamma \vert \to 0$ and small but fixed $\alpha$, both $\sqrt{-\zeta}$ and $\omega$ approach 
zero as $\sim \vert \gamma \vert$, and in the limit of  a ``particle at rest in 
a magnetic field'' the internal motion is ``frozen'', whereby
the magnitude of the spin vector (\ref{spinM}) is preserved.

For the two other configurations corresponding, for $\vert \gamma \vert > 
\alpha/2$, to the canonical direction of gyration and to
\begin{eqnarray*}
c)\quad f=f_+&,&\gamma>0\quad\;\;(\vec n \parallel \vec u)\\
d)\quad f=f_-&,&\gamma<0\quad(-\vec n \parallel \vec u),
\end{eqnarray*}
the parameters $\sqrt{-\zeta}$, $\omega$ and three-velocity $u$ behave as 
$$\sqrt{-\zeta} \simeq \alpha(1-\frac{\alpha}{2\vert \gamma \vert})^{-1},~~~
\omega \simeq \Omega \sqrt{\alpha}(1-\frac{\alpha}{2\vert 
\gamma \vert})^{-1/2},~~~u=\frac{\alpha \omega}{2\Omega},$$ 
and diverge at the {\it critical value} of three-momentum $\vert \gamma \vert = \alpha /2$. 
Interestingly, the singular behavior happens when the radius of 
gyration $R=l/2$ is equal to that in the ``free'' case. Note also that for 
the second kind of function $f_-=\sqrt{1-\sqrt{-\zeta}}$ the solution disappears
before the critical value is reached, namely, at $\sqrt{-\zeta}=1$;   
however, the difference in the values for the critical parameters is very small, and 
we neglect it for sake of simplicity. 

One should further take into account that {\it physical observables} should be defined 
according to the laboratory time scale, not to the proper time one. In particular,
the {\it physical} three-velocity $v$ and frequency of rotation $\omega_{\rm ph}$ are defined as follows:
\be{phys}
\frac{v}{c} = \frac{u}{\sqrt{1+u^2}}, ~~~ \omega_{\rm ph} = \omega \sqrt{1-(v/c)^2}
\equiv \frac{\omega}{\sqrt{1+u^2}},
\ee
so that near the critical point $\vert \gamma \vert = \alpha/2$, where
the four-velocity $u$ and proper-time frequency $\omega$ increase without limits,
the physical velocity approaches the {\it speed of light}, $v\rightarrow c$, and 
for the physical frequency of gyration one obtains 
\be{physomega}
\omega_{\rm ph} = \frac{\omega}{\sqrt{1+u^2}}\simeq \frac{\omega}{u}= \frac{\Omega}
{\vert \gamma \vert} \simeq \frac{2 \Omega}{\alpha} = \frac{2mc^2}{\hbar}.
\ee       
Thus, near the critical point the frequency of gyration remains finite and 
close to one half of the internal {\it de Broglie frequency} $\Omega_{DB}$ 
associated with the particle,   
\be{DB}
\omega_{\rm ph} = \frac{1}{2} \Omega_{DB},~~~~\hbar \Omega_{DB} = mc^2.
\ee

\begin{figure}[!ht]
\begin{center}
\includegraphics[scale=.9]{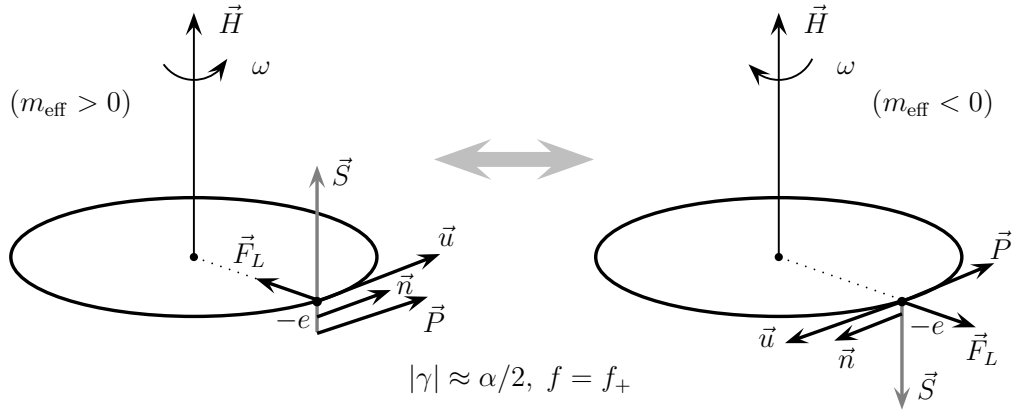}
\caption{\label{spin4} 
The rotation in a magnetic field is discontinuous at
a critical value of the three-momentum. Shown are the 
canonical direction of rotation ($m_{\rm eff}>0$)  and 
the rotation in direction opposite to the force ($m_{\rm eff}<0$).
In both cases $f=f_+$ and $\vert\gamma\vert\approx \alpha/2$.}
\end{center}
\end{figure}

In order to better understand what could happen with a particle when its 
three-momentum approaches the critical value, let us examine the opposite range for
the three-momentum, $\vert \gamma \vert < \alpha/2$. In this range, apart from
the two formerly considered configurations a) and b) with canonical direction of rotation, 
two unusual and at macro-level forbidden configurations with {\it direction of rotation opposite 
to the canonical one} can occur (for $f=f_+$ see Fig.\ref{spin4}, right panel). The particle 
is accelerated in the direction opposite to the Lorentz force acting on it, and
to explain this formally one has to prescribe a {\it negative effective 
mass} ($m_{eff}<0$) to the particle.
For such unusual configurations, the parameters $\sqrt{-\zeta},\,\omega$ and $u$ behave
in the  critical range as 
$$\sqrt{-\zeta}\simeq\alpha(\frac{\alpha}{2\vert \gamma \vert} -1)^{-1},
~~~\omega\simeq \Omega\sqrt{\alpha}(\frac{\alpha}{2\vert \gamma \vert} -1)
^{-1/2},~~~u=\frac{\alpha \omega}{2\Omega},$$
and thus diverge at the critical momentum, similarly as for the cases c) and d) considered above.
Again the corresponding {\it physical} quantities remain finite,
$v \rightarrow c$ and $\omega_{\rm ph} \rightarrow \frac{1}{2}\Omega_{DB}$.  
These ``exotic'' configurations correspond to the following conditions:  
cc) $f=f_+,~\gamma<0$, or dd) $f=f_-,~\gamma>0~$. Note, however, that for
the opposite  direction of rotation $\gamma > 0$ corresponds to $-\vec n \parallel \vec u$ 
and vice versa. Thus, near the critical value of the three-momentum $\vert \gamma \vert = \alpha/2$
the configurations c) and cc) (as well as d) and dd)) have opposite 
directions of rotation, opposite directions of unit three-vector $\vec n$ and 
of the spin vector $\vec S$ (see Fig.\ref{spin4} for $f=f_+$). 

We can thus conjecture that, when the  
three-momentum of a particle rotating in a magnetic field approaches (say, through 
radiation) the critical value $\vert \gamma \vert = \alpha/2$, its dynamics  
becomes {\it irregular}, and a {\it phase-like} transition to another rotational 
state may happen, under preservation of energy, three-momentum and magnitude of 
the spin vector, but through a flip of the directions of the spin vector, 
unit three-vector and vector of three-velocity. This very much resembles 
a quantum transition with {\it spin flop} and, perhaps, can be thought 
of as its classical counterpart. However, a lot of work has still to be done to 
comprehend such a complicated and unusual dynamics, even within the framework 
of a ``textbook'' problem of motion of a charged particle in homogeneous magnetic field.  

\section{Conclusion}

We are accustomed to consider spin as a purely quantum property of 
particles. In quantum theory, discretization of spin is 
associated with the finite dimensional irreducible representations of the 
Lorentz group. In the classical limit, this should be consistent with 
universality of the values of second Casimir invariant. As a rule, however, 
this is not supported by the equations of motion and is usually taken 
to be a constant number ``by hands''.

The same is true even for a weaker requirement of {\it preservation} of a 
particle's spin {\it in external fields} demonstrated by real elementary 
particles but having no evident counterpart in external or internal 
symmetries (contrary to preservation of mass and charge). To ensure the  
preservation at the classical level, one usually~\cite{Frenkel,BMT} needs 
to impose an additional ``Frenkel-like'' orthogonality condition on the 
solutions of the equations of motion.

In this respect, the model proposed by Staruszkiewicz is quite remarkable since 
the modulus of the spin 4-vector becomes therein 
universal, one and the same for any solution, even under the presence of 
arbitrary electromagnetic field (section 2). 
Contrary to the situation in other models (see, e.g.~\cite{Frenkel,Matiss,
BMT,Corben}), here this property is completely independent on the 
equations of motion themselves, fulfilled identically and need not to 
be postulated via imposition of an additional constraint or obtained from the 
classical dynamics via a not yet well defined quantization procedure.  

From a general point of view, the proposed deformation of the standard
action seems to be quite natural, motivated by deep symmetry and other 
physical considerations and, in a sense, unambiguous. 
The two permitted forms for the generating function found above (section 2) 
may be thought of 
as corresponding to two types of charged elementary particles of  
different mass but the 
same spin (if, instead of universal $l$ the constant $\hbar/mc$  with 
universal $\hbar$ is introduced into the action). However, at present all such 
conjectures seem to be speculative.

In the model, the original {\it four-orthogonality condition} (\ref{kF}) is always  
fulfilled ``on shell'' and coordinates the spatial and spin dynamics allowed by 
the equations of motions. In particular,  
the model demonstrates a number of quite remarkable properties 
with respect to the particle's dynamics in the ``free'' case and in the presence of 
a homogeneous magnetic field (sections 3 and 4, respectively). Some of these 
(fixed radius and indefinite alteration  of the frequency of zitterbewegung, 
spin-flop ``phase-like'' transitions) strongly 
resemble quantum phenomena, others
(cyclotron frequency shift) can help to experimentally test the model. 
Of course, the model requires deep rebuilding of the presently accepted paradigm 
when one claims to relate it to real physics.

It is also noteworthy that some properties peculiar for the model under 
consideration (plane character of the 
Zitterbewegung, magnitude of its radius equal to half a Compton wavelength etc.) 
draw it nearer to some other models, in particular to that developed by 
Rivas~\cite{Rivas,RivasAH} from rather general symmetry considerations. It 
would be interesting to find  out whether his approach and the Staruszkiewicz 
model can in fact be cooordinated and united in the starting principles as well 
as in the predicted spin dynamics.

In any case, however, mathematical beauty and simplicity of the considered 
model, its correspondence with the canonical electrodynamics in 
the spinless case, together with experimental verifiability, make it a 
nice ``toy model'' to interpret some properties of real spinning 
particles at a purely classical and visual level of consideration.

\section{Afterword}

After the paper~\cite{Our} has been accepted for publication, we 
became informed that the Staruszkiewicz model had been in fact proposed as long as 
15 years ago in a distinguished paper of Kuzenko, Lyakhovich and Segal
~\cite{Kuzenko}, on the base of elegant geometrical considerations. The authors 
had realized at least two profound ideas reproduced later in the note of 
Staruszkiewicz~\cite{Starusz}: introduction of a light-like variable 
responsible for additional spin degrees of freedom and determination of 
the action for a spinning particle from the requirement of universality 
(``strong conservation'') of the two Casimir invariants of particle's mass and 
spin. 

Specifically, they consider the ${\mathbb M}\times S^2$ geometry 
of the configuration space being a direct product of the Minkowski space and 
the two-sphere. The latter is the $SO(3,1)$-transformation space of minimal 
dimension, with (nonlinear) Lorentz action represented by fractional 
linear transformations of the component $z\in \mathbb{C}$ of the projective 
spinor $\lambda=\{1,z\}$. The $z$-component is canonically obtained 
via stereographic projection $S^2 \mapsto C_*$ of the two-sphere onto the 
Argand complex plane (for this see, e.g.,~\cite{Penrose}).   

To connect the spatial dynamics in $\mathbb{M}$ and the spin dynamics,  
the standard metric on $S^2$, 
\be{conf}
ds^2 \sim \frac{dz d\bar z}{(1+z\bar z)^2},
\ee
has been endowed with a conformal factor in which the tangent 4-vector 
${\dot x}_\mu$ to the particle's world line in $\mathbb M$ does enter. 
Finally, they arrive at the $S^2$-metric induced Lorentz invariant 
\be{Lorinv}
\Upsilon=\frac{4 \vert {\dot z} \vert^2}{(\xi \cdot {\dot x})^2}, 
\ee
where a null 4-vector 
\be{4vec}
\xi_\mu = \lambda^+ \sigma_\mu \lambda, ~~~(\xi \cdot \xi)\equiv 0
\ee
naturally comes into play. 

Together with the usual invariant $({\dot x}\cdot {\dot x})$ corresponding 
to the metric in $\mathbb M$, (\ref{Lorinv}) unambiguously predicts the form 
of the action for a free relativistic spinning particle which, after the 
elimination of the Lagrange multipliers, takes the following form (in the 
notations that we have introduced in this article):
\be{Kuzact}
S=\int L d\tau, ~~~L = -mc \sqrt{({\dot x}\cdot {\dot x})\left(1-\frac{4\Delta}
{m^2 c^2} \frac{\vert \dot z \vert}{(\xi \cdot {\dot x})}\right)} \equiv 
-mc \sqrt{({\dot x}\cdot {\dot x})\left(1-\frac{\Delta}{m^2 c^2}\sqrt{\Upsilon}\right)}, 
\ee
where the constant $\Delta$ was assumed to be $\Delta=\hbar mc\sqrt{s(s+1)}$, 
$s$ being {\it arbitrary} (at the classical level of consideration) spin of 
the particle~\footnote{This interpretation is different from that accepted 
in our article. We note that in experiments one measures only the {\it 
projection} of the spin vector and not its whole length, so the above 
identification of $\Delta$ seems doubtful.}. 

The action (\ref{Kuzact}) leads to the constant and universal values of 
both Casimir invariants and strongly resembles the Staruszkiewicz action 
(\ref{act}). To proof their equivalence, let us recall that any null 4-vector 
$k_\mu$ can be represented in the spinorial form 
\be{spinor}        
k_\mu = \omega^+ \sigma_\mu \omega.
\ee
The spinor $\omega \in {\mathbb C}^2$ is defined by (\ref{spinor}) up to a 
phase factor and, in account of the projective invariance of the principle 
$\zeta$-variable (\ref{zeta}) of Staruszkiewicz approach (see Sec. 1), 
can be reduced to the form $\omega=\{1,\tilde z\},~~\tilde z \in \mathbb C$. 
If one now makes evident identifications 
\be{identify}
\tilde z \equiv z, ~~\omega \equiv \lambda, ~~ k_\mu \equiv \xi_\mu,
\ee
then, using rules of the 2-spinor algebra, it is an easy exercise to show 
that the $\zeta$-variable (\ref{zeta}) can be expressed in the spinorial form, 
\be{zetaspin}
\zeta = l^2 \frac{(\dot{k}\cdot \dot{k})}{(k \cdot \dot  {x})^2} = -4l^2 
\frac{\vert {\dot z} \vert^2}{(\xi \cdot \dot  {x})^2}, 
\ee
and, up to a dimensional factor, coincides with the $S^2$-invariant (\ref{Lorinv}), 
\be{coincide}
-\zeta=l^2 \Upsilon. 
\ee
The action (\ref{act}), up to a sign of the additional term~\footnote{It seems that 
the possibility of two opposite signs revealed in the above presented paper, 
see (\ref{solfunc}), has been overlooked in both preceding approaches.}, also 
takes the form (\ref{Kuzact}) of the ${\mathbb M} \times S^2$ action. 

Thus, the Staruszkiewich model is indeed equivalent to that proposed earlier 
in~\cite{Kuzenko}. In the latter paper, the authors even accomplished 
a thorough Hamiltonian analysis and carried out canonical quantization 
of the free model. As for the classical part, in the free case the 
plane character and fixed radius of Zitterbewegung have been already 
established in~\cite{Kuzenko}. The only difference with the above presented 
results (Sec. 3) is in the frequency of particle's revolution which in 
~\cite{Kuzenko} was obtained arbitrary but constant. We think that this 
fact is due to the therein used time gauge, just as it takes place in our 
study of free motion under the special choice of time scale (\ref{Tscale}). 
Nonetheless, in the physical laboratory time scale 
Zitterbewegung frequency turns out to be capable of variation in an arbitrary manner. 

It is noteworthy to mention that the Kuzenko-Lyakhovich-Segal (KLS) model 
has been later generalized to the supersymmetrical case~\cite{SuperGen} and 
to the Anti-de-Sitter background geometry~\cite{AdSGen}. In the paper
~\cite{UnivGen} some additional possibilities to construct the action for 
a (both massive and massless) spinning particle in ${\mathbb M} \times S^2$ 
exploiting spinor structure of the two-sphere have been proposed, and the 
coupling with external 
electromagnetic and gravitational field formally considered. However, contrary 
to the original paper~\cite{Kuzenko} and to the Staruszkiewicz approach, the 
constructions developed in~\cite{UnivGen} require the introduction of new 
parameters with vague physical sense and, in general, seem rather cumbrous and 
ambigous. 

After the afore-presented short comparative analysis of the KLS-model and 
Staruszkiewich approach it is noteworthy to underline that nearly all 
the results obtained above (and in our paper~\cite{Our}) are new in the 
respect when an external electromagnetic field is present. This concerns to 
the universality (``strong conservation'') of the mass-spin invariants, to 
the ``on-shell'' orthogonality (\ref{kF}) of the null four-vector and 
four-force vector and the exact results related to the plane motion of 
a spinning particle in constant magnetic field (corrections to 
cyclotron frequency, possible spin-flop transitions etc.). As to the free 
case, our results mainly reproduce those obtained in~\cite{Kuzenko} 
(distinctions in the frequency of Zitterbewegung and in 
interpretation of the spin value were mentioned above and reqiure 
additional analysis). On the other hand, the two different approaches 
and the results obtained here go ``hand by hand'' and provide essential 
support to the Kuzenko-Lyakhovich-Segal-Staruszkiewich model demonstrating 
its really exceptional character. 
%\vspace{0.5cm}
\\[2mm]
\textbf{Acknowledgment:} The authors wish to thank an anonymous referee for valuable
remarks.

\end{document}